\documentclass[pra,twocolumn,preprintnumbers,amsmath,amssymb,superscriptaddress,showpacs,longbibliography]{revtex4-1}
\usepackage{graphicx}
\usepackage{amsmath}
\usepackage{graphics}
\usepackage{amssymb}
\usepackage{amsfonts}
\usepackage{dsfont}
\usepackage{natbib}
\usepackage{hyperref}
\usepackage{enumitem}
\usepackage{physics}
\usepackage{color}
\usepackage[mathscr]{eucal}

\allowdisplaybreaks

\newcommand{\beq}{\begin{equation}}
\newcommand{\eeq}{\end{equation}}
\newcommand{\bea}{\begin{eqnarray}}
\newcommand{\eea}{\end{eqnarray}}

\newcommand{\opH}{\mathbf{H}}
\newcommand{\cv}[1]{\left(#1\right)}
\newcommand{\cvb}[1]{\left[#1\right]}
\newcommand{\cvc}[1]{\left\{#1\right\}}
\newcommand{\cvv}[1]{\left\vert#1\right\vert}

\newcommand{\ave}[2]{\left\langle #2\right\rangle_{#1}}
\newcommand{\aveb}[1]{\left\langle #1\right\rangle}
\newcommand{\iden}{\openone}
\newcommand{\opV}{\mathbf{V}}

\newcommand{\opS}{\hat{\sigma}}

\newcommand{\opP}{\mathbf{P}}

\newcommand{\prob}{\mathbb{P}}

\begin{document}

\title{On relationship between subjecting the qubit to dynamical decoupling and to a sequence of projective measurements}
\author{Fattah Sakuldee}
\email{sakuldee@ifpan.edu.pl}
\affiliation{Institute of Physics, Polish Academy of Sciences, al.~Lotnik{\'o}w 32/46, PL 02-668 Warsaw, Poland}
\author{{\L}ukasz Cywi\'{n}ski}\email{lcyw@ifpan.edu.pl}
\affiliation{Institute of Physics, Polish Academy of Sciences, al.~Lotnik{\'o}w 32/46, PL 02-668 Warsaw, Poland}

\begin{abstract}
We consider a qubit coupled to another system (its environment), and discuss the relationship between the effects of subjecting the qubit to either a dynamical decoupling sequence of unitary operations, or a sequence of projective measurements. 
We give a formal statement concerning equivalence of a sequence of coherent operations on a qubit, precisely operations from a minimal set $\left\{\openone_Q,\hat{\sigma}_x\right\}$, and a sequence of projective measurements of $\opS_x$ observable. 
Using it we show that when the qubit is subjected to $n$ such successive projective measurements at certain times, the expectation value of the {\it last} measurement can be expressed as a linear combination of expectation values of $\opS_x$ observed after subjecting the qubit to dynamical decoupling sequences of $\pi$ pulses, with $k\leq n$ of them applied at subsets of these times. 
Performing a sequence of measurements on the qubit gives then access to the same properties of the environment and qubit-environment coupling that are affecting the coherence observed in a dynamical decoupling experiment. Analysing the latter has been widely used to characterize the environmental dynamics (perform so-called noise spectroscopy), so our result shows how the results obtained with dynamical decoupling based protocols are related to those that can be obtained just by performing multiple measurements on the qubit. 
We also discuss in more detail the application of the general result to the case of the qubit undergoing pure dephasing, and outline possible extensions to higher-dimensional (a qudit or multiple qubits) systems.
\end{abstract}

\date{\today}
		
\maketitle

\section{Introduction}
Unitary operations on an open quantum system are commonly employed in information manipulation since they preserve the purity of the state. 
They are used for control of quantum information \cite{Schumacher2010,Bennett2018,Heinosaari2012} encoded initially in the system, and the open nature of the system is most often treated as a nuisance - a source of decoherence \cite{Schlosshauer_book}. Good definitions of the quantum system ($S$) and its environment ($E$) are in fact based on the possibility of exerting unitary control: $S$ can be subjected to any desired unitary operation, while $E$ can only be manipulated in a very limited (if any at all) way. Here we also assume that $S$ can also be subjected to projective measurements, while its $E$ is inaccessible to both unitary operations and measurements. 
The only way then to learn anything about $E$ is {\it through the manipulations and measurements on $S$}. Gaining information about the dynamics of $E$  could be then used to devise system control protocols that perform desired tasks while being more resilient to decoherence \cite{Gordon_PRL08,Biercuk_Nature09,Uys_PRL09,Glaser_EPJD15,Poggiali_PRX18}. Furthermore, characterization of a microscopic (but still not directly controllable) environments of certain qubits, e.g.~nuclear spins of molecules localized close to a nitrogen-vacancy center qubit \cite{Degen_RMP17,Staudacher_Science13,DeVience_NN15,Haberle_NN15,Lovchinsky_Science16,Pfender_NC19,Cujia_Nature19}, is of interest in itself, as it allows to use the coherent control and readout of a small quantum system to gain insight into physics of another, larger, quantum system.

In recent years the most popular qubit-based environment characterization method that based on applying a sequence of short unitary operations, e.g.~$\pi$ pulses corresponding to application of $\hat{\sigma}_k$ ($k\! =\! x$, $y$, $z$) operators, at a chosen set of times, followed by a single measurement of the qubit \cite{Degen_RMP17,Szankowski_JPCM17,Almog_JPB11,Biercuk_JPB11,Bylander_NP11,Alvarez_PRL11,Kotler_Nature11,Medford_PRL12,Staudacher_Science13,Muhonen_NN14,Romach_PRL15,Malinowski_PRL17}, has been intensively developed. Such a procedure is known as {\it dynamical decoupling} (DD) \cite{Viola_PRA98,Suter_RMP16}, as it was originally devised to protect the qubit from decoherence by decoupling it from $E$, but in the context relevant here the goal is to decouple the qubit (or qubits) from all the environmental noise, {\it except for noise at certain frequencies}, and thus turn the quantum system into a spectrometer of this noise \cite{deSousa_TAP09,Cywinski_PRB08,Biercuk_JPB11,Degen_RMP17,Szankowski_JPCM17}.    
In the case of pure dephasing of the qubit, and $E$ being either a source of external classical noise, or (possibly quantum) Gaussian noise, the relation between the DD signals and the properties of the environmental dynamics is well-established \cite{Szankowski_JPCM17,Degen_RMP17,Norris_PRL16,Paz_PRA17}. By an appropriate choice \cite{Alvarez_PRL11,Szankowski_PRA18} of DD sequences one can reconstruct the power spectral density of Gaussian noise, and characterization of polyspectra of non-Gaussian noise is also possible, although more challenging \cite{Norris_PRL16,Sung_NC19,Ramon_PRB19}. Note that this basic setup for $E$ characterization is similar to the one used in quantum metrology \cite{Giovanetti_Science04}, where one employs protocols consisting of unitary operations on the multi-qubit system followed by a single measurement \cite{Huelga_PRL97,Kolodynski_NJP13,Szankowski_PRA14}, or of periods of unitary evolution interlaced with measurements used for error correction of the state of the qubits \cite{Demkowicz_PRL14,Dur_PRL14,Kessler_PRL14}.

Projective measurements, on the other hand, are of entanglement breaking character, and never allow for continuity of the correlation between $S$ and $E$ in the further steps of the protocol \cite{Ruski2003,Pollock2018,StochasticTraj}. 
This class of operations is most often employed for characterization of the system \cite{DiLorenzo2013,Chantasri2018a}. It is possible to exert some degree of control over the system by subjecting it to a sequence of measurements, often also involving post-selection \cite{Pechen_PRA06,Hill_PRA08,Ashab_PRA10,Blok_NP14,Muhonen_PRB18}.
While multiple measurements on a qubit are known to allow for characterization and changing the state of $E$ that is static during its interaction with the qubit \cite{Klauser_PRB06,Stepanenko_PRL06,Giedke_PRA06,Sakuldee_PRA19}, only quite recently the possibility of performing such characterization of {\it dynamics} of $E$ by performing only measurements on $S$ has attracted more attention \cite{Liu_NJP10,Fink_PRL13,Bechtold_PRL16,Muller_SR16,Wang_PRL19,Do_NJP19,Muller2020,Pfender_NC19,Cujia_Nature19,Bethke_arXiv19,Sakuldee_classical_PRA20}. Building on earlier results \cite{Fink_PRL13}, we have recently established a close connection between the DD-based, and multiple measurement-based noise spectroscopy, in the case of pure dephasing due to external classical noise \cite{Sakuldee_classical_PRA20}. However, the model in which a quantum $E$ is replaced by a source of classical noise, while apparently often accounting well for observations in the pure dephasing case \cite{Degen_RMP17,Szankowski_JPCM17,Biercuk_JPB11,Bylander_NP11,Alvarez_PRL11,Kotler_Nature11,Medford_PRL12,Staudacher_Science13,Muhonen_NN14,Romach_PRL15,Malinowski_PRL17}, is obviously a drastic approximation to the microscopic description  of the open quantum system dynamics.
The issue of applicability of multiple measurements to characterization of a {\it quantum} $E$ coupled to the controlled system, is thus a subject of present interest, see e.g. a recent work on reconstruction of arbitrary higher order environmental correlation functions from sequential measurements on the qubit \cite{Wang_PRL19}.

There are several pieces of evidence showing that the two above-mentioned classes of operations can provide similar characteristics of the environment, when several sequences of measurements are together taken into account. Examples include observation that positive operator valued measures (POVMs) can be explained in terms of projective measurements \cite{Oszmaniec2017,Oszmaniec2018}; derivation of noisy quantum channel decoding efficiency bound via the expansion over projective sequential measurements \cite{Lloyd2011}; projective measurement(-preparations) reconstruction of non-Markovian dynamics over limited controls \cite{Milz2018}, and explanation of non-Markovian control in terms of alternative formalism of quantum stochastic process constructed from a set of measurement-preparation pairs \cite{StochasticTraj}. 

In this work we show that a close relationship between dynamical decoupling of $Q$ from $E$ and the  protocols based on multiple measurements on $Q$, is in fact very general: it holds on the operational level without making any assumption on the initial state of the total system, the$Q$-$E$ coupling, and the quantum or effectively classical nature of environmental dynamics. 
Specifically, we consider control over a qubit generated from a minimal set of two operations $\cvc{\iden_Q,\opS_x}$, and we show how arbitrary sequences of such operations interlaced with unitary evolutions of the composite ($Q+E$) system, can be expressed as linear combinations of operations in which these basic unitary operations on $Q$ are replaced by measurements of its $\opS_x$ (and vice versa). The main result of experimental significance is establishment of relation between signals obtained using a class of protocols based on multiple measurements, and measurements of coherence of a qubit subjected to a DD sequence of $\pi$ pulses (i.e.~$\opS_x$ operations). 
We expect this result to contribute to the recently ongoing theoretical efforts aimed at understanding what characteristics of the quantum environment one can obtain from multiple measurements on the qubit \cite{Ma_PRA18,Gefen_PRA18,Wang_PRL19}, and at extending the DD-based noise spectroscopy paradigm to the case of general qubit-environment coupling \cite{Paz2019}. 
We also discuss the relevance of this relation for noise spectroscopy in the pure dephasing case, and outline the generalization to higher-dimensional systems.

The paper is organized in the following way. The basic setup of the considered composite system,  the mathematical framework,  and conventions for coherent control and sequential measurement protocol are given in Section \ref{sec:frame}. In Section \ref{sec:rel_op} we derive the main formal result: we express the operation done on the composite system by a sequence of measurements as a linear combination over DD unitary evolutions followed by a single measurement, and, conversely, we express the operation done by a sequence of $\opS_x$ unitaries as a linear combination of projections interlaced with unitary evolutions. 
Then, in Section \ref{sec:rel_sig} we focus on the application of this result to the observables most easily accessible in the experiment: decoherence signal after a DD sequence and expectation the last measurement in a sequence of projections. We also discuss some features specific to the often-encountered case of pure dephasing of the qubit, as this is the case for which most of DD-based noise spectroscopy theory (for a recent exception see \cite{Paz2019}) was developed. 
Some possible generalizations to measurements along multiple axes and higher-dimensional system case are sketched in Section \ref{sec:some_gen}, while in the concluding Sec.~\ref{sec:discussion}, apart from summarizing the main results,  we put them in the context of recent works on characterization of dynamics of open quantum systems using the process tensor \cite{Pollock2018,Modi2012}.

\section{Framework}\label{sec:frame}
\subsection{Open system subjected to interventions}
Our main focus here is a qubit $Q$ coupled to an environment $E$. The Hamiltonian of the composite $Q+E$ system is 
\begin{equation*}
\opH_C = \opH_{Q}\otimes\mathds{1}_{E} +\mathds{1}_{Q}\otimes\opH_{E} + \sum_{k=x,y,z}\hat{\sigma}_{k}\otimes\opV_{k} \,\, ,
\end{equation*}
where $ \opH_{Q(E)}$ are Hamiltonians of $Q(E),$ $\hat{\sigma}_{k}$ are Pauli operators of the qubit, and $\opV_k$ are environmental operators. Between times $t_{k-1}$ and $t_{k}$ the evolution of the system is either generated this Hamiltonian, with $\mathbf{U}^{C}_{k} \! =\!  \exp[-i\opH_C (t_k-t_{k-1})]$, or by the Hamiltonian of the environment only: $\mathbf{U}^{E}_{k} \! =\!  \exp[-i\opH_E (t_k-t_{k-1})]$. The latter occurs when the $Q$-$E$ coupling and the qubit Hamiltonian is turned off for a finite time. Such a temporary vanishing of the $Q$-$E$ interaction often occurs naturally in setups in which measurements on $Q$ at some of $t_{k}$ times are considered: these measurements could be destructive, or their execution might require changing the system in  such a way that $Q$ and $E$ become decoupled. 

We consider now various types of interventions at times $t_k$, all of them being operations local on $Q$, as only the latter is considered to be directly accessible. The goal is to choose these operations, the intervention times, and the pattern of types of unitary evolutions between these times ($\mathcal{U}^{C}_k$ or $\mathcal{U}^{E}_k$), in a way allowing for recovery of certain properties of both $E$ (its Hamiltonian and initial state), and the $Q$-$E$ coupling, from measurements on the final state of $Q$. 
More generally, one might want to look  at patterns in correlations between the results of this measurement, and the input state of the composite system. 
Both quantum metrology (where typically more than one qubit forms the system, and $E$ is replaced by a real number $x$ on which $\mathbf{U}^C_k$ depends), and $E$ characterization with $\pi$ pulses applied to the qubit are examples. In the former, most accurate estimation of $x$ is the goal. In the latter, it is harder to state in general terms what is the quantity of interest. From a body of work on DD-based noise spectroscopy \cite{Degen_RMP17,Szankowski_JPCM17}, we know that for a qubit initialized in a superposition of pointer states, periodic application of a sequence of unitary operations on the qubit, possibly interlaced with operations that decouple it from $E$ (see \cite{Fink_PRL13,Laraoui_NC13,Gefen_PRA18,Sakuldee_classical_PRA20} for examples), followed by a measurement of qubit's coherence, gives access to physically interesting properties of dynamics caused by $\opH_C$ and $\opH_E$. We would like now to formulate a more general framework of controlling the evolution of the composite system by a series of interventions concerning $Q$ only, with the DD-based scheme being one particular example. 

\subsection{Operational formalism}
In order to treat various types of interventions on equal footing, we will use an operational formalism, in which both unitary evolution and measurement on a system described by state operator $\rho$ are identified with superoperators acting on $\rho$. For example, a unitary evolution $\mathbf{U}$ corresponds to operation $\mathcal{U}$, the action of which is $\mathcal{U}[\rho] = \mathbf{U}\hat{\rho}\mathbf{U}^{\dagger}$.

Aside from formal convenience, this formalism allows for transparent separation of discussion concerning the {\it operation} on the total system, caused by our interventions and its intrinsic dynamics, and consideration of input state and the final measurement that constitutes the ``signal'' obtained from a given procedure. We will show that relations between distinct classes of interventions can be formulated by considering only the {\it operational} layer: they hold for all the input states. 

We consider a qubit $Q,$ with an arbitrary environment $E$, undergoing unitary evolution $\mathcal{U}$ acting on a state $\rho\in\mathcal{S}\cv{\mathcal{H}_Q\otimes\mathcal{H}_E}=\cvc{\rho\in\mathcal{B}\cv{\mathcal{H}_Q\otimes\mathcal{H}_E}: \tr{\rho}=1,~\rho \geq 0,~ \rho=\rho^\dagger}$. Let us define a sequence of $n$ time steps $\cv{t_n,t_{n-1},\ldots,t_1}$ with $t_n>t_{n-1}>\ldots>t_1$, and write $\mathcal{U}_k=\mathcal{U}\cv{t_k,t_{k-1}}$ for all $k=1,\ldots,n$. As discussed previously, they can correspond either to an evolution in presence of $Q$-$E$ interaction, or they can correspond to evolution of $E$ only. 
We call an operation on $\mathcal{B}\cv{\mathcal{H}}$ an operator thereon equipped with Hilbert-Schmidt algebra that obeys the conditions of complete positivity \cite{Heinosaari2012,UnitalSeparation}, and we call the operation local on the subsystem $Q$ if $\mathcal{B}\cv{\mathcal{H}_Q}\otimes\mathbf{B},$ where $\mathbf{B}$ is an environment operator, is closed under the operation. 

\begin{figure}[tb]
\begin{flushleft}
a.)
	\includegraphics[width=\linewidth]{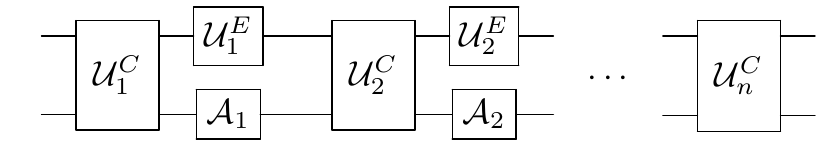}
	\hfill
b.)
	\includegraphics[width=\linewidth]{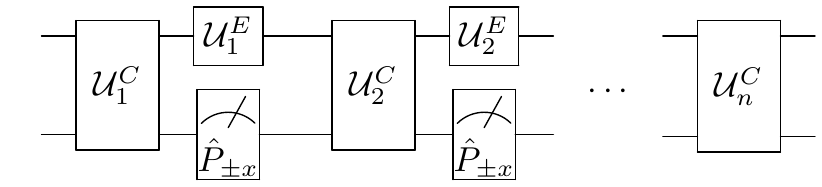}
	\caption{Schematics of sequences of (a) coherent operations, and (b) sequential measurements. The modules above represent the operations on the composite states, while the input state and output state(observable) are arbitrary. The upper rail of the scheme describes the evolution of the environment, while the lower rail of the scheme describes the operations done on the qubit. $\mathcal{U}^{C}_{k}$ operations correspond to joint evolution of the qubit and the environment caused by their interaction, while $\mathcal{U}^{E}_{k}$ correspond to evolution of $E$ uncoupled from the qubit.}\label{fig:schematic}
\end{flushleft}
\end{figure}

We consider two types of local operations: coherent ones and projective ones. The first preserve purity of the state in the qubit subsystem, i.e.~$\tr{\cvb{\text{tr}_E\{\mathcal{A}\cvb{\rho}\}}^2}=\tr{\cvb{\text{tr}_E\{\rho\}}^2}$ where $\text{tr}_E$ is a partial trace over environment degrees of freedom. For  a qubit, we will focus on a restricted class consisting of {\it idle} operation $\mathcal{I}\cvb{\rho}=\rho$, and {\it single-axis $\pi$ pulse} operation $\mathcal{X}\cvb{\rho} = \cv{\opS_x\otimes\iden_E}\rho\cv{\opS_x\otimes\iden_E}$ where $\opS_x$ is a Pauli $X$ operator. In this language the evolution of the composite system modulated by a sequence of length $n-1$ of  coherent operations on $Q$ is given by
	\begin{equation}
		\mathcal{U}^{\mathcal{A}}_{s_{n-1},\ldots,s_1}=\mathcal{U}_n\circ\mathcal{A}_{s_{n-1}}\circ\ldots\circ\mathcal{U}_2\circ\mathcal{A}_{s_1}\circ\mathcal{U}_1\label{eq:coherent_concatenation}
	\end{equation}
where $\mathcal{A}_{s_k}=\displaystyle\left\{\begin{array}{lr}\mathcal{I},~s_k=\cv{i}\\ \mathcal{X},~s_k=\cv{x}	\end{array}\right.$ and $s_k\in\cvc{\cv{i},\cv{x}}$ denotes the sequence of idle $\cv{i}$ and echo $\cv{x}$ operation at time steps labeled by $k.$ For example, a spin echo sequence \cite{deSousa_TAP09} corresponds to the operation $\mathcal{U}_2\circ\mathcal{X}\circ\mathcal{U}_1$ with $\mathcal{U}_2 \! = \! \mathcal{U}_1 \! = \! \mathcal{U}^{C}(\tau)$ describing evolution of the composite system for time $\tau$, while a two pulse Carr-Purcell (CP) sequence \cite{Haeberlen} corresponds the operation $\mathcal{U}_3\circ\mathcal{X}\circ\mathcal{U}_2\circ\mathcal{X}\circ\mathcal{U}_1$ with identical generators of the evolutions and their durations being $\tau,2\tau$, and $\tau$, consecutively. A schematic representation of a sequence of local coherent $\mathcal{X}$ operations, with evolution of either interacting $Q$-$E$ composite system, or $E$ only occuring between these operations, is shown in Fig.~\ref{fig:schematic}a.

Another class of operations is of the projective type. In accordance with our focus on $\mathcal{X}$ operation above we consider the measurement in $X$ basis done by projections $\opP_{\pm}$ onto the $\ket{+}$ and $\ket{-}$ eigenstates of $\opS_x,$ i.e. $\opS_x\ket{\pm}=\pm\ket{\pm},$ associated with a measurement outcome $m=\pm 1$. The corresponding operation is $\mathcal{P}_m\cvb{\rho} = \cv{\opP_{m}\otimes\iden_E}\rho\cv{\opP_{m}\otimes\iden_E}.$ It corresponds to an entanglement breaking channel \cite{Ruski2003}, since a correlated state will become separable after the measurement, i.e. the purity of the composite state will be increasing. In a concatenated form, a sequence of $n$ measurements interlaced with $n$ evolutions is given by 
	\begin{equation}
		\mathcal{P}_{m_n,\ldots,m_1}=\mathcal{P}_{m_{n}}\circ\mathcal{U}_n\circ\mathcal{P}_{m_{n-1}}\circ\ldots\circ\mathcal{U}_2\circ\mathcal{P}_{m_1}\circ\mathcal{U}_1\label{eq:measurement_concatenation}
	\end{equation}
where $m_k\in\cvc{+1,-1}$ denotes a measurement result at time step $k,$ and the sequence of unitary evolutions $\mathcal{U}_n,\ldots,\mathcal{U}_1$ is the same as in Eq.~(\ref{eq:coherent_concatenation}), see Fig.~\ref{fig:schematic}b.

\section{Relation between Coherent Operations and Sequential Measurements}\label{sec:rel_op}
\begin{figure}[tb]
\begin{flushleft}
	\includegraphics[width=\linewidth]{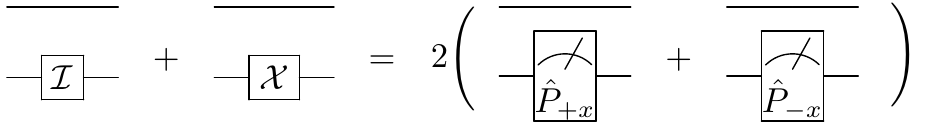}
	\caption{Schematics of relation of the building blocks for the equivalence of two types of manipulation: left hand side is a combination of coherent operations on the qubit and the right hand side is a combination of projective measurements on it.}\label{fig:block}
\end{flushleft}
\end{figure}

Since $\opP_m$ and $\opS_x$ commute, one may describe operation from one class as combinations of the operations from the other class. In particular we have a relation $\mathcal{X}=2\cv{\mathcal{P}_+ + \mathcal{P}_-}-\mathcal{I}$ (see Fig.~\ref{fig:block} for the schematics.), or conversely $\mathcal{P}_m=\dfrac{1}{4}\cv{\mathcal{I}+\mathcal{X}+m\mathcal{D}_X}$ where $\mathcal{D}_X\cvb{\rho}=\cv{\opS_x\otimes\iden_E}\rho + \rho\cv{\opS_x\otimes\iden_E}$. Thus, it follows that
	\begin{align*}
		\mathcal{P}_{m_n,\ldots,m_1} &= \dfrac{1}{4^n}\left[\cv{\mathcal{I} + \mathcal{X} + m_n\mathcal{D}_X}\circ\mathcal{U}_{n}\right.\\
		 &\hspace{1.2cm} \left.\circ\cdots\circ\cv{\mathcal{I} + \mathcal{X} + m_1\mathcal{D}_X}\circ\mathcal{U}_{1}\right].
	\end{align*}
	As one can see, on the operation level, the 
	direct expansion of $\mathcal{P}_{m_n,\ldots,m_1}$ will contain $\mathcal{D}_X$ in the sequences. However, we can avoid this feature if we consider an operation $\mathcal{O}_n$ defined by
	\begin{equation*}
	\mathcal{O}_n\cv{m_n} = \sum_{m_{n-1},\ldots,m_1}\mathcal{P}_{m_n,\ldots,m_1} \,\, , 
	\end{equation*}
	which corresponds to making $n-1$ {\it non-selective} measurements \cite{vonNeumann1955,Kalev2013,He2019,Pechen_PRA06} (or measurements without outcome evaluation) at times $t_1,\ldots,t_{n-1}$, followed by projection on $m_n$ at time $t_n.$ It is easy to see now that
	\begin{equation}
			\mathcal{O}_n\cv{m_n} = \dfrac{1}{2^{n-1}}\cvb{\mathcal{P}_{m_n}\circ\sum_{s_{n-1},\ldots,s_1}\mathcal{U}^{\mathcal{A}}_{s_{n-1},\ldots,s_1}}  \,\, .
		 \label{eq:Meas-in-DD}
	\end{equation}
	The operation $\mathcal{O}_n\cv{m_n}$ is effectively a composition of all possible $2^{n-1}$ sequences of coherent operations applied to the qubit at times $t_{1},\ldots,t_{n-1}$, defined in Eq.~(\ref{eq:coherent_concatenation}), followed by a measurement giving $m_n$ result at $t_n$. We remark that 
	$\sum_m m\mathcal{C} \! = \!0$ for any outcome-independent operation $\mathcal{C}.$ The relation (\ref{eq:Meas-in-DD}) means that a measurement $\mathcal{P}_{m_n}$ at time $t_n$ preceded by earlier $n-1$ non-selective measurements, represented by $\mathcal{O}_n\cv{m_n}$, transforms the whole composite system in the same way as a procedure in which we perform unitary transformations $\mathcal{U}^{\mathcal{A}}_{s_{n-1},\ldots,s_1}$ with probabilities $1/2^{n-1}$, and follow each of them by $\mathcal{P}_{m_n}$ measurement. For example, for a sequence of two measurements performed a times $t_1$ and $t_2$, we have
	\begin{align*}
		\mathcal{O}_2\cv{m_2} &= \sum_{m_1}\mathcal{P}_{m_2}\circ\mathcal{U}_2\circ\mathcal{P}_{m_1}\circ\mathcal{U}_1\\
			&=  \dfrac{1}{2}\cv{\mathcal{P}_{m_2}\circ\mathcal{U}_2\circ\mathcal{U}_1 + \mathcal{P}_{m_2}\circ\mathcal{U}_2\circ\mathcal{X}\circ\mathcal{U}_1}
	\end{align*}
	showing that  the state after two measurements can be written as a convex combination of states obtained by making a single measurement at time $t_2$ with no control pulse, and with $\pi$ pulse at time $t_1.$ This relation is the basis of the classical environmental noise characterization scheme  by two single-shot measurements described in \cite{Fink_PRL13}.
	
For the converse relation, it suffices to consider only the case of $\cv{\cv{x},\cv{x},\ldots,\cv{x}}$, because an insertion of an idle operation between two unitary evolutions can be absorbed in a redefinition of unitary evolution $\mathcal{U}_k\circ\mathcal{I}\circ\mathcal{U}_{k-1}\rightarrow\mathcal{U}_{k-1}$ with shifting of indices $k+1\mapsto k,$ mapping the original operation sequence to a shorter one containing only $\mathcal{X}$ operations. For example, the sequence $\cv{\cv{x},\cv{i},\cv{x}}$ can be written as $\cv{\cv{x},\cv{x}}$ by redefinition $\mathcal{U}_3\circ\mathcal{I}\circ\mathcal{U}_{2}\rightarrow\mathcal{U}_{2}$ from the original sequence. For $n-1$ operations $\mathcal{X}$ interlaced in the sequence with $n$ unitary evolutions, we have
	\begin{equation}
		\begin{split}
			\mathcal{U}^{\mathcal{A}}_{\cv{x},\ldots,\cv{x}} &= \sum_{k=0}^{n-1}\cv{-1}^{n-1-k}2^{k}~\mathcal{U}'_{\cv{t_\ell}_k}\\
			&\hspace{1cm}\circ\sum_{\cv{t_\ell}_k\in\mathfrak{T}_{n-1}}\sum_{m_k,\ldots,m_1}\mathcal{P}'_{m_k,\ldots,m_1}
		\end{split}\label{eq:DD-in-meas}
	\end{equation}
	where $\mathfrak{T}_{n-1}$ is a set of all possible sub-sequences $\cv{t_\ell}_k$ of length $k,$ for $k=1,\ldots,n-1,$ of the sequence of operation times $\cv{t_{n-1},\ldots,t_1},$ $\mathcal{U}'_{\cv{t_\ell}_k}$ is a composition of all the unitary evolutions after the last measurement in the subsequence $\cv{t_\ell}_k$, and $\mathcal{P}'_{m_k,\ldots,m_1}=\prod_{\ell=1}^k\cv{\mathcal{P}_{m_\ell}\circ\mathcal{U}'_\ell}$ with $\mathcal{U}'_\ell=\mathcal{U}_\ell\circ\cdots\circ\mathcal{U}_{\ell-1}$ being the composition of all unitary evolutions between measurement time steps $t_{\ell-1}$ and $t_\ell$ in the subsequence $\cv{t_\ell}_k$, at which the measurements $\mathcal{P}_{m_\ell}$ are evaluated. For example, the operation that corresponds to spin echo sequence is
	\begin{equation*}
	\mathcal{U}_2 \circ\mathcal{X}\circ\mathcal{U}_1 = -\mathcal{U}_2\circ\mathcal{U}_1 + 2\cv{\mathcal{U}_2\circ\mathcal{P}_+\circ\mathcal{U}_1 + \mathcal{U}_2\circ\mathcal{P}_-\circ\mathcal{U}_1} \,\, .  
	\end{equation*}
	Thus, letting the composite system evolve for time $t_1$, applying $\mathcal{X}$ to the qubit, and then letting the system evolve for time $t_2-t_1$, can be written as a linear combination of three other operations: evolution for time $t_2$, evolution for $t_1$ followed by projection on $\ket{+}$ state and subsequent evolution for $t_2-t_1$, and an analogous operation with projection on $\ket{-}$ state at $t_1$. For the evolution interrupted by $2$ pulses, we have
	\begin{align*}
		\mathcal{U}_3&\circ\mathcal{X}\circ\mathcal{U}_2\circ\mathcal{X}\circ\mathcal{U}_1\\
		&= \mathcal{U}_3\circ\mathcal{U}_2\circ\mathcal{U}_1\\
		& - 2\cv{\mathcal{U}_3\circ\mathcal{U}_2\circ\mathcal{P}_+\circ\mathcal{U}_1 + \mathcal{U}_3\circ\mathcal{U}_2\circ\mathcal{P}_-\circ\mathcal{U}_1}\\
		& - 2\cv{\mathcal{U}_3\circ\mathcal{P}_+\circ\mathcal{U}_2\circ\mathcal{U}_1 + \mathcal{U}_3\circ\mathcal{P}_-\circ\mathcal{U}_2\circ\mathcal{U}_1}\\
		&+ 4\left(\mathcal{U}_3\circ\mathcal{P}_+\circ\mathcal{U}_2\circ\mathcal{P}_+\circ\mathcal{U}_1 + \mathcal{U}_3\circ\mathcal{P}_-\circ\mathcal{U}_2\circ\mathcal{P}_-\circ\mathcal{U}_1\right.\\
		& + \left.\mathcal{U}_3\circ\mathcal{P}_+\circ\mathcal{U}_2\circ\mathcal{P}_-\circ\mathcal{U}_1 + \mathcal{U}_3\circ\mathcal{P}_-\circ\mathcal{U}_2\circ\mathcal{P}_+\circ\mathcal{U}_1\right).
	\end{align*}
	Note that the right-hand sides of the two above equations are {\it not} convex combinations of operations, and consequently one cannot perform an experiment in which the operations effected by $\pi$ rotations on the qubit are performed with qubit measurements only. In the next Section we discuss the observable consequences of these relations.

\section{Decoherence Signals and Repreparations for Pure-Dephasing}\label{sec:rel_sig}
Our main results, Eqs. \eqref{eq:Meas-in-DD}-\eqref{eq:DD-in-meas}, show that the relation between sequences of coherent operations and projective measurements on the qubit can be expressed at the level of operations on the composite system. However, it is more practical to consider an expectation of a particular observable of the qubit, which can be studies in experiments. 

\subsection{Relation between Dynamical Decoupling Induced Decoherence Signal and Probabilities of Sequential Measurements}
\begin{figure}[tb]
\begin{flushleft}
a.)
	\includegraphics[width=\linewidth]{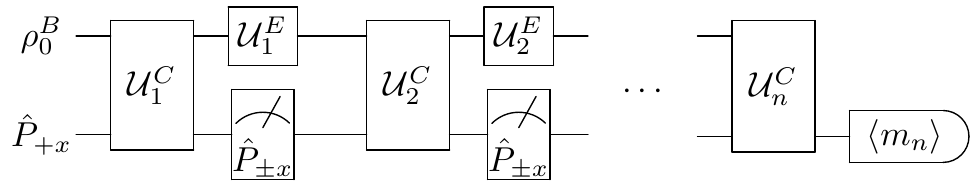}
	\hfill
b.)
	\includegraphics[width=\linewidth]{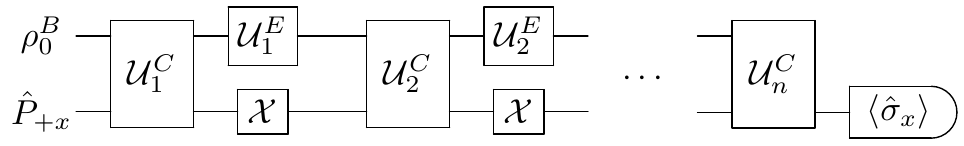}
	\hfill
c.)
	\includegraphics[width=\linewidth]{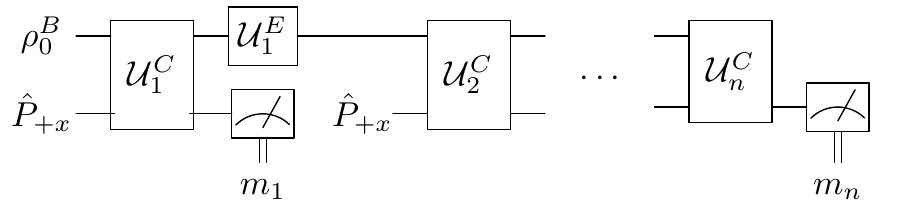}
	\caption{Schemes of three protocols leading to signals observable on the qubit discussed in Section \ref{sec:rel_sig}:
	(a.) average of final measurement result from the collection of measurements without re-preparation, (b.) decoherence signal induced by dynamical decoupling process, and (c.) correlation of the collection of measurements with re-preparation of the qubit state. The signal from protocol (a) can be expressed as a linear combination of signals from (b), and vice-versa, for any form of qubit-environment coupling. The signals from (b) and (c) are equal one to another for qubit exposed to pure dephasing due to coupling of its $\hat{\sigma}_z$ operator to the environment.}\label{fig:observable}
\end{flushleft}
\end{figure}

Let us apply Eqs. \eqref{eq:Meas-in-DD}-\eqref{eq:DD-in-meas} to a system initially in the state $\rho=\opP_+\otimes\rho_E,$ and consider the expectation value of qubit's $\opS_x.$ In a sequential measurement protocol, we calculate an expectation of the $n^{th}$ measurement $O_n\cv{t_n,\ldots,t_1} = \displaystyle\sum_{m_n}m_n\tr{\mathcal{O}_n\cv{m_n}\cvb{\rho}}$, see Fig.~\ref{fig:observable}a.
It can be written as $\displaystyle O_n\cv{t_n,\ldots,t_1} = \sum_{m_n,\ldots,m_1}m_n\prob\cv{m_n,\ldots,m_1}$ where $\prob\cv{m_n,\ldots,m_1} = \tr{\mathcal{P}_{m_n,\ldots,m_1}\cvb{\rho}}$ is a probability of a sequence of results.

On the other hand, we consider an expectation value of $\hat{\sigma}_x$ of the qubit at time $t_n$, evaluated after the qubit was subjected to a sequence of coherent operations $\mathcal{U}^{\mathcal{A}}_{s_{n-1},\ldots,s_1}$:
	\begin{equation*}
		W_{s_{n-1},\ldots,s_1}\cv{t_n,\ldots,t_1}:=\tr{\cv{\opS_x\otimes\iden_E}\mathcal{U}^{\mathcal{A}}_{s_{n-1},\ldots,s_1}\cvb{\rho}} \,\, ,
\end{equation*}	
see Fig.~\ref{fig:observable}b. This function measures coherence between eigenstates of $\hat{\sigma}_z$ for a qubit initialized in their superposition, and subjected to a DD sequence of $\pi$ pulses about the $x$ axis. From Eq.~\eqref{eq:Meas-in-DD}, we obtain a simple relation between the results of the two experiments
	\begin{equation}
		O_n\cv{t_n,\ldots,t_1} = \dfrac{1}{2^{n-1}}\sum_{s_{n-1},\ldots,s_1} W_{s_{n-1},\ldots,s_1}\cv{t_n,\ldots,t_1} \,\, .\label{eq:On-in-Wn}
	\end{equation}
This is a generalization of a relation obtained in \cite{Sakuldee_classical_PRA20} for a qubit experiencing pure dephasing due to external classical noise. However, here we {\it do not} assume anything about the form of qubit-environment coupling, and the environment is treated quantum mechanically. 

These relations, show that the collection of expectations of $\opS_x,$ given by all measurement sub-sequences of length $k<n,$ 
reveals the same characteristics of the environment as those which are commonly extracted from a collection of decoherence signals obtained in DD experiments with $n-1$ or less pulses.

Let us consider the converse relation, as we have done on the operation level. Let us define an operation in which the first $k-1$ measurements are done at times specified by the subsequence $\cv{t_\ell}_{k-1}$ of $\cv{t_{n-1},\ldots,t_1}$, and the $k^{th}$ measurement is done at time $t_n,$
	\begin{equation*}
		\mathcal{O}_k\cv{m_k;t_n\oplus\cv{t_\ell}_{k-1}} = \displaystyle\sum_{m_{k-1},\ldots,m_1} \!\!\!\!\! \mathcal{P}_{m_k}\circ~\mathcal{U}'_{\cv{t_\ell}_{k-1}}\circ\mathcal{P}'_{m_{k-1},\ldots,m_1} \,\, ,
	\end{equation*}
and the corresponding expectation value of $k^{th}$ measurement:
	\begin{equation*}
	 	O_k\cvb{t_n\oplus\cv{t_\ell}_{k-1}}=\displaystyle\sum_{m_k,\ldots,m_1} \!\!\!\!\! m_k\tr{\mathcal{O}_k\cv{m_k;t_n\oplus\cv{t_\ell}_{k-1}}\cvb{\rho}}.
	\end{equation*} 
In this language Eq.~\eqref{eq:DD-in-meas} leads to 
	\begin{equation}
		\begin{split}
			W_{\cv{x},\ldots,\cv{x}}\cv{t_n\ldots,t_1}& \\
			&\hspace{-2.2cm}=\sum_{k=0}^{n-1}\cv{-1}^{n-1-k}2^{k}\sum_{\cv{t_\ell}_{k-1}\in\mathfrak{T}_{n-1}}O_k\cvb{t_n\oplus\cv{t_\ell}_{k-1}}.
		\end{split}\label{eq:Wn-in-Ok}
	\end{equation}
For example, the decoherence signals from a spin echo protocol (the decoherence signal as a function of $t_2$ with a pulse applied at $t_1$) can be written as a composition of results of two measurement protocols
	\begin{equation}
		W_{\cv{x}}\cv{t_2,t_1} = 2O_2\cv{t_2,t_1} - O_1\cv{t_2} \,\, ,\label{eq:W1-in-O}
	\end{equation}
	where $O_{2}(t_2,t_1)$ is an expectation value of $\opS_x$ measured at time $t_2$ when a previous measurement of this observable was done at time $t_{1}$, while $O_{1}(t_2)$  is an expectation value of $\opS_x$ measured at time $t_2$ that was not preceded by another measurement. 
	The converse relation, obtained from Eq.~(\ref{eq:On-in-Wn}), is simply
	\begin{equation}
		O_2\cv{t_2,t_1} = \dfrac{1}{2}\cv{W_{\cv{i}}\cv{t_2,t_1} + W_{\cv{x}}\cv{t_2,t_1}} \,\, ,\label{eq:O2-in-W1}
	\end{equation}
	where $W_{\cv{x}}\cv{t_2,t_1}$ ($W_{\cv{i}}\cv{t_2,t_1}$) is the expectation value of $\opS_x$ measured at time $t_2$ when a $\pi$ pulse(identity) operation was applied at time $t_{1},$ so that in fact the $t_1$ argument in $W_{\cv{i}}\cv{t_2,t_1}$ is spurious.

Another example is a three measurement protocol for which the expectation value of the  last measurement is 
		\begin{equation}
			\begin{split}
				O_3\cv{t_3,t_2,t_1} &= \dfrac{1}{4}\left(W_{\cv{i},\cv{i}}\cv{t_3,t_2,t_1} + W_{\cv{x},\cv{i}}\cv{t_3,t_2,t_1}\right.\\
				&\hspace{0.5cm}\left. + W_{\cv{i},\cv{x}}\cv{t_3,t_2,t_1} + W_{\cv{x},\cv{x}}\cv{t_3,t_2,t_1}\right)\label{eq:O3-in-W2}
			\end{split}
		\end{equation}
	with interventions done at $t_2$ and $t_1$, and coherence is measured at $t_3.$  On the other hand, for two pulse Carr-Purcell sequence with inter-pulse delays given by $\tau,2\tau,\tau$, we have
		\begin{equation}
			\begin{split}
				W_{\text{CP-2}}\cv{4\tau} &= O_1\cv{4\tau} - 2\cvb{O_2\cv{4\tau,\tau}+ O_2\cv{4\tau,3\tau}} \\
				&\hspace{0.5cm}+ 4O_3\cv{4\tau,3\tau,\tau},\label{eq:W_CP-in-O}
			\end{split}
		\end{equation}
	where measurements are done at times given as arguments of $O_{k}$ functions.

\subsection{Relation between measurement protocols with and without re-preparation in case of pure dephasing}
Above we have considered the sequence of projective measurements without any re-preparation of the states. However, in practice one may desire an insertion of re-preparation of a particular state. This is the case when the measurement is destructive, and the qubit needs to be re-prepared in a fresh state such as $\opP_{+}$. The measurement operation will then be followed by a re-preparation operation $\mathcal{R}\cvb{\rho}=\opP_{+}\otimes\text{tr}_Q\{\rho\}$. The measurement protocol with re-preparation will be given by
	\begin{equation}
	 	\mathcal{P}^{\mathcal{R}}_{m'_n,\ldots,m'_1}=\mathcal{P}_{m'_{n}}\circ\mathcal{U}_n\circ\mathcal{R}\circ\mathcal{P}_{m'_{n-1}}\circ\ldots\circ\mathcal{U}_2\circ\mathcal{R}\circ\mathcal{P}_{m'_1}\circ\mathcal{U}_1,\label{eq:measurement_withre-prepare_concatenation}
	 \end{equation} 
	where the prime symbol indicates the measurement results in this protocol. This is schematically illustrated in Fig.~\ref{fig:observable}c.

Let us focus now on the often-encountered case of pure dephasing evolution, in which one can find a basis for $Q$  that consists of pointer states unperturbed by coupling to $E$ \cite{Zurek_PRD82,Zurek_RMP03,Schlosshauer_book}, and only superposition of these states are subjected to  a dephasing. The Hamiltonian is then of the form $\opH=a_z\opS_z\otimes\opV_z + a_1\iden_Q\otimes\opV_1$ where $\opS_z$ and $\iden_Q$ are Pauli $Z$ and identity operator on the qubit, $\opV_{z(1)}$ is an operator acting on $\mathcal{H}_E,$ and $a_{z(1)}$ is a (possibly time dependent) real number. 
This Hamiltonian describes a dominant decoherence mechanism for a wide class of qubits \cite{Degen_RMP17,Szankowski_JPCM17,Almog_JPB11,Biercuk_JPB11,Bylander_NP11,Alvarez_PRL11,Kotler_Nature11,Medford_PRL12,Staudacher_Science13,Muhonen_NN14,Romach_PRL15,Malinowski_PRL17,Roszak_PRA06}.  The unitary evolution is given then by a conjugation 
with  $\mathbf{U}=\op{\uparrow}\otimes\mathbf{U}_\uparrow + \op{\downarrow}\otimes\mathbf{U}_\downarrow$ where $\op{\uparrow}$ and $\op{\downarrow}$ are projections onto eigenstates of $\opS_z.$

We observe the following relabelling relation between measurement sequences Eqs. \eqref{eq:measurement_concatenation} and \eqref{eq:measurement_withre-prepare_concatenation} at the level of probabilities:
	\begin{equation}
		\prob_{\mathcal{R}}\cv{m'_n,m'_{n-1},\ldots,m'_1}\equiv\prob\cv{m_{n}\cdot m_{n-1},\ldots,m_2\cdot m_1,m_1},\label{eq:M_with-vs-without}
	\end{equation}
where $\prob_{\mathcal{R}}$ is a probability of obtaining a sequence of results in the protocol with re-preparation. 
The input state $\opP_{\pm}$ will be relabelled as $\opP_{\mp}$ if the previous measurement result is $\opP_{-}$, and not be relabelled otherwise. Note that this relabelling simply corresponds to a change of {\it assignment convention} of the measurement results to the measurement sequences. A derivation of this relation is given in Appendix~\ref{appen:re-prep_derive}.

The previously considered expectation value $O_n$ corresponds then to a  {\it correlation} of all the measurement results in the re-preparation case:
	\begin{align*}
		O_n\cv{t_n,\ldots,t_1} &= \sum_{m_n,\ldots,m_1}m_n\prob\cv{m_n,\ldots,m_1}\\
			&= \sum_{m'_n,\ldots,m'_1}\cv{\prod_{k=1}^nm'_k}\prob_{\mathcal{R}}\cv{m'_n,\ldots,m'_1} \,\, .
	\end{align*}
Consequently, the relations between decoherence signals induced by sequences of $\pi$ pulses, and expectations over measurement sequences, can also be applied to the protocol with re-preparation when the expectation $O_n$ is replaced by the correlation of all the measurements in the sequence. Schematic representations of the measurement protocol without re-preparation, dynamical decoupling,  and measurement protocol with re-preparation, are given in Fig.~\ref{fig:observable}. 

In an experiment with two measurements, we have $O_2\cv{t_2,t_1}=\langle\opS_x\cv{t_2}\opS_x\cv{t_1}\rangle$, and using Eq. \eqref{eq:O2-in-W1} we arrive at the result obtained in \cite{Fink_PRL13} for $E$ being a source of classical noise. Here we have shown, without making any assumption about the nature of $E,$ that a correlation of $n$ measurements of $\opS_x$, each followed by a re-initialization of the qubit in $\ket{+}$ state, is related to measurements of coherence of qubits subjected to dynamical decoupling according to Eq.~\ref{eq:On-in-Wn}. This generalizes the relationships  derived in \cite{Sakuldee_classical_PRA20} for classical environmental noise, to quantum environments. 

\section{Some Possible Generalisations}\label{sec:some_gen}

\subsection{Two-axes manipulation}\label{par:two-axes}
Apart from the minimal control algebra $\cvc{\iden_Q,\opS_x}$, pulses about other axes can be also considered. For example, in the sequences of coherent operations one may replace some of $\mathcal{A}_k$ by $\mathcal{Y}$ operations in the protocols without re-preparation. In the corresponding measurement protocol we will use then the relations $\mathcal{Y}=2\cv{\mathcal{P}_+^Y + \mathcal{P}_-^Y}-\mathcal{I}$ and $\mathcal{P}^Y_{m_Y}=\dfrac{1}{4}\cv{\mathcal{I} + \mathcal{Y} -im_Y\mathcal{D}_Y}$, where the outcomes $m_Y$  have $\pm i$ values assigned in order to distinguish them from $\mathcal{P}^X_{m_X}$ measurement, and the elementary operations are defined as for the $X$ axis. One can see that the coherent sequences, as well as the sequential measurements with an additional measurement axis $Y$, can be considered as an intertwining of sequences from the control sets $\cvc{\iden_Q,\opS_x}$ and $\cvc{\iden_Q,\opS_y}$. For instance, a sequence $\mathcal{A}^Y\circ\mathcal{U}\circ\mathcal{A}^X,$ in which $\mathcal{A}^{X(Y)}$ is a $\pi$ pulse with respect to $X(Y)$ axis, can be related to $\mathcal{P}^Y_{m_Y}\circ\mathcal{U}\circ\mathcal{P}^X_{m_X}$ in a similar fashion as in Eqs. \eqref{eq:Meas-in-DD}-\eqref{eq:DD-in-meas} without additional difficulty.  This agrees with the results in Ref.~\cite{Wang_PRL19}, where the higher order bath correlations (which can be obtained by pulse sequences in principle, see e.g.~\cite{Norris_PRL16}) are extracted from measurements along multiple axes. 

However, in the pure dephasing case and for the protocol with re-preparations, the relabeling procedure becomes now more complicated, since some sequences will contain subsequent operations along distinct axes, e.g.~$\mathcal{Y}\circ\mathcal{U}\circ\mathcal{X}$. The re-preparation operation $\mathbb{R}$ will map the four possible outcomes $\opP_\pm$ and $\opP_\pm^Y$ into $\opP_+.$ The relabeling will be possible both in a single-axis and two-axis cases, because $\cvc{+1,-1,+i,-i}$ is still closed under multiplication, and one can construct the relabeling convention similar to the one in Eq.~\eqref{eq:M_with-vs-without}. Furthermore, $\opP_\pm$ together with $\opP_\pm^Y$ can be considered as a tomography basis for the subsystem evolution of the qubit, where the set of all positive value operators $\cvc{\opP_\pm,\opP_\pm^Y}$ will no longer fully orthogonal, but symmetric and informationally complete (SIC-POVM) \cite{Renes2004}. Hence it would be interesting to consider other choice of measurement for the expansions of coherent operations, e.g.~a tetrahedral basis in the Bloch sphere \cite{Renes2004}. 

\subsection{Multi-qubit manipulation}\label{par:multi-qubit}
\begin{figure*}[tb]
\begin{flushleft}
	\includegraphics[width=\linewidth]{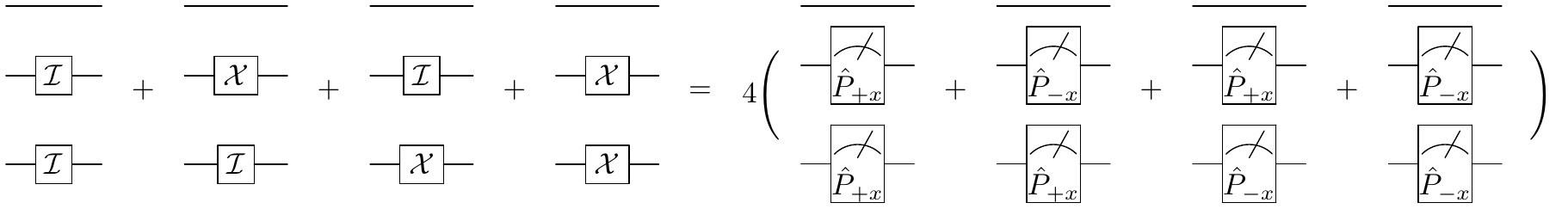}
	\caption{Schematic of relation of the building blocks for the equivalence of two types of manipulation in two qubits case.}\label{fig:block-2}
\end{flushleft}
\end{figure*}
We have focused so far on the case of a single qubit, but the situation in which multiple qubits are coupled to $E$, and coherent operations and measurements on all of them are used to characterize it, has been a subject of recent works on multi-qubit generalizations of DD-based $E$ characterization protocols \cite{Szankowski_PRA16,Paz_NJP16,Paz_PRA17,Muller_SR18,Krzywda_NJP19}.
Let us show now that in this case we can also find a relation between the evolution of the system due to $\mathcal{X}$ operations and $\mathcal{P}_{\pm}$ measurements on the qubits.
This is easily seen from the relation $\mathcal{I}+\mathcal{X}^{(j)}=2\cv{\mathcal{P}_+^{(j)}+\mathcal{P}_-^{(j)}},$ where the superscript denotes the qubit number. Note that we consider now the situation in which only local $\mathcal{X}^{(j)}$ operations are applied, and we do not consider nonlocal operations such as two-qubit swaps, taken into account in some multi-qubit generalization of DD-based protocols \cite{Paz_PRA17}. The global operation is then given by 
	\begin{equation}
		\prod_{j=1}^n\cv{\mathcal{I}+\mathcal{X}^{(j)}}=2^n\prod_{j=1}^n \cv{\mathcal{P}_+^{(j)}+\mathcal{P}_-^{(j)}} \label{eq:n-qubits}
	\end{equation}
	and the expansion on both sides will lead to the basic relation for single step intervention on $n$ qubits. For example, with two qubits we obtain
	\begin{equation}
		\begin{split}
			\mathcal{I}&+\mathcal{X}^{(1)}+\mathcal{X}^{(2)}+\mathcal{X}^{(1)}\circ\mathcal{X}^{(2)}\\
			&= 4\bigg(\mathcal{P}_+^{(1)}\circ\mathcal{P}_+^{(2)}+\mathcal{P}_-^{(1)}\circ\mathcal{P}_+^{(2)}\\
			&\hspace{1cm}+\mathcal{P}_+^{(1)}\circ\mathcal{P}_-^{(2)}+\mathcal{P}_-^{(1)}\circ\mathcal{P}_-^{(2)}\bigg).
		\end{split} \label{eq:2-qubits}
	\end{equation}
	The circuit representation of this relation is given in Fig.~\ref{fig:block-2}. The concatenation form and interlacing with system-environment interaction evolution $\mathcal{U}^{I}_k$ will follow in the similar fashion as in the single qubit case. From this observation, we see that a local $\pi-$pulse manipulation protocol on multi-qubit system can be related to the statistics of measurements on multiple qubits. Importantly, since we have considered only local coherent operations, in the above relation we need only {\it local} measurements, as each $\mathcal{P}^{(j)}_{\pm}$ is a measurement operation concerning only the $j$-th qubit.

\subsection{Multi-dimensional system}\label{par:multi-dim}
For a general finite system, the expansion of coherent operations in terms of identity operations and measurement operations can be implemented in several ways, depending on the control algebra. One of the simplest examples is a sequential shifting protocol \cite{StochasticTraj,Jena2019} over $d-$dimensional system with the control set $\cvc{\iden,\mathbf{S}_1,\ldots,\mathbf{S}_{d-1}}$ generated by $\mathbf{S}_k=\mathbf{g}^k$, with a shifting generator
	\begin{equation}
		\mathbf{g}=\left(\begin{array}{cccccc}
			0&1&0& & &0\\
			0&0&1&\ldots& &0\\
			 &\vdots& &\ddots& &\vdots\\
			 & & & &0&1\\
			1&0&0&\ldots&0&0
		\end{array}\right).\label{eq:g_matrix}
	\end{equation}
Eigenprojections $\cvc{\opP_0,\opP_1,\ldots,\opP_{d-1}}$ of the matrix $\mathbf{g}$ define the corresponding measurement axes, while eigenvalues belong to the set $\cvc{m_0=1,m_1,\ldots,m_{d-1}};$ all possible $d^{th}$ roots of $1$, namely the solutions of $z^d=1$, will be assigned as measurement values to all projections. From the structure of the measurement outcomes, one can deduce that $\displaystyle 1+ \sum_{j=1}^{d-1}m_j = 0,$ $\cvv{m_j}=1$ for all $j$, and $\cvc{m_j}_{j=1}^{d-1}$ is closed under multiplication and complex conjugation. Now we write $\mathcal{S}_k\cvb{\rho}=\mathbf{S}_k\rho\mathbf{S}_k^\dagger,$ $\mathbf{S}_k=\displaystyle\sum_{i=0}^{d-1}m^k_i\opP_{i},$ $\mathcal{P}_i\cvb{\rho}=\opP_i\rho\opP_i$ and $\mathcal{Q}_{ij}\cvb{\rho}=\opP_i\rho\opP_j.$  It follows that 
	\begin{equation}
		\mathcal{I} + \sum_{k=1}^{d-1}\mathcal{S}_k = d\sum_{i=0}^{d-1}\mathcal{P}_i +\sum_{i=0}^{d-1}\sum_{j\neq i}\cvb{1 + \sum_{k=1}^{d-1}\cv{m_i\overline{m}_j}^k}\mathcal{Q}_{ij}\label{eq:gen_shift_map}
	\end{equation}
	where $\overline{m}$ is a complex conjugate of $m.$ The element $m_i\overline{m}_j$ is also a root of unity. If all roots except $1$ are primitive, e.g. $d$ is a prime number, the set $\cvc{m_j}_{j=0}^{d-1}$ will be equal to $\cvc{m_j^k}_{k=0}^{d-1}$ for any $m_j,~j\neq0$ \cite{lang_1966}. Hence the terms in the double summation vanish, and we obtain an analogue  of $\mathcal{I}+\mathcal{X}=2\cv{\mathcal{P}_++\mathcal{P}_-}$ as
	\begin{equation}
		\mathcal{I} + \sum_{k=1}^{d-1}\mathcal{S}_k = d\sum_{i=0}^{d-1}\mathcal{P}_i,\label{eq:gen_shift_map_prime}
	\end{equation}
	or simply speaking the overall shifting procedure can be reproduced by sequential measurements in the same basis. For the case with more than one non-primitive root $m_j$, the structure may be folded into sub-cycle of shorter lengths for the order of its division. For example, with $d=4$ we find that $\cvc{1,-1,i,-i}^2=\cvc{-1,1}$, which is the set of measurement outcomes for two dimensions, and the elements in the double sum still vanish in this case. In addition, from Eq.\eqref{eq:gen_shift_map_prime}, contrary to $\mathcal{I}+\mathcal{X}=2\cv{\mathcal{P}_++\mathcal{P}_-},$ for dimension $d>2$ one cannot fully express the effect from single shifting operation, e.g. $\mathcal{S}_k$ for some $k,$ in terms of only measurement operations and identity operations, but operations of all orders (e.g. $\mathcal{S}_k$ for every $k,$) need to be taken into account.

\section{Discussion and Conclusion}\label{sec:discussion}
We have presented two main results concerning the composite system of a qubit (Q) and its environment (E). The first is a formal statement on the equivalence of effects from sequence of coherent local operations (for the minimal set of control $\cvc{\iden_S,\opS_x}$) on the qubit, and sequential projective measurements on it. We have shown that the operation effected on the composite system  in one of these ways, can be expressed as a linear combination of operations from the other class. This holds for any initial state of the whole system (including correlated qubit-environment states), and for any form of qubit-environment interaction. 

The second result, following from the first one, is the relation between observables obtained in two kinds of experiments: one involving $n$ projective measurements (at times $t_1,\ldots,t_n$) of $\opS_x$ on the qubit, and the other involving application of $k\! < \! n$ rotations by $\pi$ about the $x$ axis ($\opS_x$ operations) at times forming a subset of times of first $n-1$ of measurements, followed by measurement of $\opS_x$ at the $t_n$ time. 
For an initially uncorrelated Q-E state, the expectation value of the last measurement in the first experiment can be expressed as a linear combination of expectations values of $\opS_x$ in the second experiment. We have also shown the converse result: the decoherence signal measured after subjecting the qubit to a sequence of $\pi$ pulses  can be replicated by using observables obtained from multiple sequences of  measurements on the qubit.

In the commonly encountered case of pure dephasing of the qubit (with $E$ coupling only to $\sigma_z$ operator of the qubit), such sequences of $\pi$ pulses about the $x$ axis lead to frequency-selective dynamical decoupling (DD) of $Q$ from its $E$, which has been widely used to characterize the environmental dynamics \cite{Degen_RMP17,Szankowski_JPCM17,Almog_JPB11,Biercuk_JPB11,Bylander_NP11,Alvarez_PRL11,Kotler_Nature11,Medford_PRL12,Staudacher_Science13,Muhonen_NN14,Romach_PRL15,Malinowski_PRL17}. We have thus shown how all the results for DD-based spectroscopy of qubits undergoing pure dephasing (with all the $\pi$ pulses about the same axis), can be recovered with protocols in which the qubit is subjected solely to multiple measurements, and this holds for a general environment described quantum mechanically.

We have also discussed the protocol considered in \cite{Fink_PRL13,Sakuldee_classical_PRA20}, in which the qubit is re-initialized in a chosen state after each measurements, and correlations between results of multiple measurements are considered. In the pure dephasing case its results are related to the linear combination of DD signals in the same way as the result of the above-discussed multiple-measurement protocol. As a consequence, the noise spectroscopy protocols considered in \cite{Fink_PRL13} and \cite{Sakuldee_classical_PRA20} for the case of the $E$ being a source of classical noise, can be also employed for a truly quantum environment.

A natural framework for the discussion of these results is provided by the concept of a process tensor introduced in Refs.~\cite{Pollock2018,Modi2012}. It is the mapping from the sequence of operations $\Phi_1,\ldots,\Phi_{n-1}$ (e.g. coherent operations or measurement sequence) to the final state at time $t_n$. The dynamics of the quantum system and its initial state can be treated as a single unknown entity that one wants to study, by subjecting it to arbitrary quantum operations at a set of times $t_1,\ldots,t_{n-1}$. In the setup considered in the paper, the system consists of a subsystem that we can control and measure (a qubit, or a higher-dimensional system as discussed in Sections \ref{par:multi-qubit} and \ref{par:multi-dim}), and a subsystem that is not accessible directly (an environment). The goal then is to characterize the process tensor of the composite system by performing operations on the controllable subsystem only, and then, after taking into account additional information on the initial state of this subsystem, to characterize the influence of the environment on this subsystem. 

The first main result of the paper shows how the mappings corresponding to unitary operations on the qubit are related to those corresponding to measurements, i.e.~it concerns the general structure of the process tensor of such a composite system. The projective measurements on the qubit break the entanglement between qubit and environment, so it seems then that some information about the character of the composite system dynamics will be lost when we consider a single sequence of projections. 
However, as we have shown here, by combining the sequences of projections in a specific way, given in Eqs.~\eqref{eq:Meas-in-DD}, we can recover the effect that one obtains from a sequence of unitary interventions. This means that, at least for the minimal set of control operations considered here, using coherent operations on the qubit does not provide any theoretical advantage (practically it might be, of course, more efficient to implement) in characterization of the process tensor of the open quantum system. 

Compared to the case of the qubit, extending the formulation to the case of higher dimensional systems is challenging, with a system of $n$ qubits subjected to local $\pi$ pulses being an exception. For qubits we have discussed a specific case of shifting protocols, but for arbitrary control sets the formulation of an analogous relation between a class of protocols based on coherent operations on a subsystem, and on measurements on this subsystem, should be considered on a case by case basis. Further work in this direction, and establishing a more general connection between the two modes of manipulation of open quantum systems, remains open for further investigation.

\section*{Acknowledgements}
We thank Jan Krzywda, Damian Kwiatkowski, and {Piotr} Sza\'{n}kowski for discussions. 
We are grateful to Felix Pollock  for a careful reading of the manuscript and valuable suggestions. 
This work is supported by funds of Polish National Science Center (NCN), Grant no.~2015/19/B/ST3/03152.

\appendix
\section{Relation between protocols with and without re-initialization of the qubit}\label{appen:re-prep_derive}
In this section we will consider in more detail the evolution of composite system generated by $\opH=a_z\opS_z\otimes\opV_z + a_1\iden_Q\otimes\opV_1$ in the main text. 
We know that the state after $(k-1)^{th}$ measurement will be of the form $\dfrac{1}{2}\cv{\iden+p_k\opS_x}\otimes\rho^B_{k-1}$, where $p_k=\pm 1$ according to the measurement outcome of the $(k-1)^{th}$ measurement. After an evolution $\mathcal{U}_k$ followed by measurement in the state $\dfrac{1}{2}\cv{\iden + m_k\opS_x}$, the unnormalised state will be 
			\begin{equation}
				\dfrac{1}{2}\cv{\iden + m_k\opS_x}\otimes\rho^B_{k} = \dfrac{1}{2}\cv{\iden + m_k\opS_x}\otimes\cv{\mathcal{K}_{m_k,p_k}\cvb{\rho^B_{k-1}}}
			\end{equation}
			where $\mathcal{K}_{m_k,p_k}\cvb{\rho^B}=\mathbf{K}_{m_k,p_k}\rho^B\mathbf{K}^\dagger_{m_k,p_k},$ $\mathbf{K}_{m_k,p_k}=\dfrac{1}{2}\cv{\mathbf{U}_\uparrow\cv{\tau_k}+p_km_k\mathbf{U}_\downarrow\cv{\tau_k}}$ and $\mathbf{U}_{\uparrow(\downarrow)}\cv{\tau_k}=e^{-i\tau_k\cv{a_1\opV_1\pm a_z\opV_z}}$ with  duration of the evolution given by $\tau_k.$ 
			
From the the environment point of view, as can be deduced from the the reduced map $\mathcal{K}$, it can be said that the effect on the environment from the measurement does not truly depend on the outcome state $\ket{\pm}$, but on the difference in sign between the outcome and the incoming state; or in other words one can write $\mathcal{K}_{m_k,p_k}=\mathcal{K}_{m_k\cdot m_{k-1},+}.$ 
This holds for the pure dephasing case, since the average dynamical map is unital, and the Bloch ball can be separated into two subspaces concerning $\cv{1Z}$ and $\cv{XY}$ plane. Consequently, the transformation $\cv{x,y}\mapsto\cv{-x,-y}$ while the $z$ is kept, can be done without disturbing the structure of the dynamics \cite{UnitalSeparation}.
			
Using this notation, in addition to the measurement sequence we can consider $\cv{p_1,\ldots,p_n}$ as a sequence of preparations, and then the probability of getting measurement sequence $\cv{m_1,\ldots,m_n}$ given a sequence of preparation $\cv{p_1,\ldots,p_n}$ reads 
			\begin{equation}
				\prob\cv{m_n,\ldots,m_1\vert~ p_n,\ldots,p_1}=\tr{\cv{\prod_{k=n}^1\mathcal{K}_{m_k,p_k}}\cvb{\rho^B_0}}. \label{eq:main_prob}
			\end{equation}
The protocol without re-preparation (the scheme considered in Sections III and IVA) can be described by the set of parameters $p_1=+1$ and $p_k=m_{k-1}$ for $k>1$, while the protocol with re-preparation in $\ket{+}$ (considered in Sec.~IVB) will be denoted as $p_k=+1$ for all $k\geq 1.$ From the observations in the previous paragraph one can see that 
			\begin{equation}
				\begin{split}
					\prob&\cv{m_1,\ldots,m_n\vert +1,\ldots,+1}\\ &= \prob\cv{m_1,m_1\cdot m_2,\ldots,m_n\cdot m_{n-1}\vert +1,m_1,\ldots,m_{n-1}}
				\end{split}
			\end{equation}
	so that the probabilities from the protocol with re-preparation can be bijectively mapped to that from the protocol with re-preparation in only $\ket{+}.$

From a statistical point of view, it is clear that a moment or measurement correlation observed from the procedure with re-preparation, can be obtained from the statistics of the protocol without re-preparation. For instance, an $n$ measurement correlation in the case with re-preparation, $\ave{\mathcal{R}}{\displaystyle\prod_{k=1}^nm_k}$, can be reproduced from the expectation of the last measurement result from the protocol without re-preparation:
		\begin{align}
			\ave{\mathcal{R}}{\prod_{k=1}^nm_k} &= \sum_{m_k} \cv{\prod_{k=1}^nm_k}\prob\cv{m_1,\ldots,m_n\vert +,\ldots,+}\label{eq:m_corr_def}\\
			&=\sum_{m_k} m_n\cv{\prod_{k=1}^{n-1}m_k^2}\nonumber\\
			&\hspace{0.5cm}\times \prob\cv{m_1,m_2,\ldots,m_n\vert +,m_1,\ldots,m_{n-1}}\nonumber\\
			\ave{\mathcal{R}}{\prod_{k=1}^nm_k} &=\aveb{m_n}=O_n\cv{t_n,\ldots,t_1} \label{eq:prep-noprep_equiv}
		\end{align}
where $\aveb{A}=\sum_{m_k}A~\prob\cv{m_n,\ldots,m_1\vert~ p_n,\ldots,p_1}.$ 

We remark again that this property holds due to two factors: (i) the manipulated system is a qubit, so the choice of measured and prepared states is limited to $\cvc{+1,-1},$ and they can be related easily, and (ii) we consider the pure dephasing Hamiltonian, so the plane subspace $\cv{XY}$ and invariant subspace $\cv{1Z}$ will be evolve separately. In order to intuitively understand  the origin of this relation, the basic idea is that in the protocol without re-preparation, the probability to get a particular measurement result at any time step depends on the previous measurement results. Consequently, the signal obtained from the last measurement result will contain the characteristics of the whole measurement sequence. On the other hand, in the protocol with re-preparation this situation cannot occur, so the experimenter needs to collect all the measurement results to obtain the same statistics as in the previous case.

\bibliographystyle{apsrev4-1}
\bibliography{Simulating_Coherent_Operations_PRA_revision.bbl}

\end{document}